\documentclass[11pt]{llncs}
\usepackage{yhmath}
\usepackage{url}
\usepackage{amssymb,wasysym}
\usepackage{graphicx}
\usepackage{wrapfig}
\usepackage{subfig}
\usepackage{color}
\usepackage{multirow}
\usepackage{epstopdf}
\usepackage[colorlinks=true]{hyperref}

\usepackage[usenames,dvipsnames]{xcolor,colortbl}

\def\rknn{\mbox{R$k$NN}}
\def\r1nn{\mbox{R$1$NN}}
\def\knn{\mbox{$k$NN}}
\def\knng{\mbox{$k$-NNG}}
\def\ksyg{\mbox{$k$-SYG}}
\def\1syg{\mbox{$1$-SYG}}

\newcommand{\etal}{\emph{et~al.}}

\newcommand{\eg}{\emph{e.g.}}

\graphicspath{{images/}}

\urldef{\mailsa}\path||
\urldef{\mailsb}\path||
\urldef{\mailsc}\path||
\newcommand{\keywords}[1]{\par\addvspace\baselineskip
\noindent\keywordname\enspace\ignorespaces#1}

\mainmatter  

\title{Kinetic Reverse $k$-Nearest Neighbor Problem~\footnote{ This work was partially supported by a British Columbia Graduate Student Fellowship and by NSERC discovery grants.}}

\author{\small Zahed Rahmati, Valerie King, and Sue Whitesides}

\institute{\small Department of Computer Science, University of Victoria\\ {\tt \{rahmati, val, sue\}@uvic.ca}}

\toctitle{Kinetic Reverse $k$-Nearest Neighbor Problem}
\tocauthor{Z. Rahmati~\etal}

\titlerunning{Kinetic Reverse $k$-Nearest Neighbor Problem}
\authorrunning{Z. Rahmati~\etal}

\begin{document}

\maketitle

\begin{abstract}
This paper provides the first solution to the kinetic reverse $k$-nearest neighbor (\rknn) problem in $\mathbb{R}^d$, which is defined as follows: Given a set $P$ of $n$ moving points in arbitrary but fixed dimension $d$, an integer $k$, and a query point $q\notin P$ at any time $t$, report all the points $p\in P$ for which $q$ is one of the $k$-nearest neighbors of $p$.


\end{abstract}
\setcounter{page}{1}
\section{Introduction}
The \textit{reverse $k$-nearest neighbor} (\rknn) problem is a popular variant of the $k$-nearest neighbor (\knn) problem and asks for the influence of a query point on a point set. Unlike the \knn~problem, the exact number of reverse $k$-nearest neighbors of a query point is not known in advancem, but as we prove in this paper the number is upper-bounded by $O(k)$. The \rknn~problem is formally defined as follows: Given a set $P$ of $n$ points in $\mathbb{R}^d$, an integer $k$, $1\leq k\leq n-1$, and a query point $q\notin P$, find the set $\rknn(q)$ of all $p$ in $P$ for which $q$ is one of $k$-nearest neighbors of $p$. Thus $\rknn(q)=\{p\in P:~|pq|\leq |pp_k|\}$, where $|.|$ denotes Euclidean distance, and $p_k$ is the $k^{th}$ nearest neighbor of $p$ among the points in $P$. The \textit{kinetic \rknn} problem is to answer \rknn~queries on a set $P$ of moving points, where the trajectory of each point $p\in P$ is a function of time. Here, we assume the trajectories are polynomial functions of maximum degree bounded by some constant $s$.
\vspace{-5pt}
\paragraph{\textbf{Related work.}} The reverse $k$-nearest neighbor problem was first posed by Korn and Muthukrishnan~\cite{Korn:2000:ISB:342009.335415}  in the database community, and then considered extensively in this community due to its many applications, \eg, decision support systems, profile-based marketing, traffic networks, business location planning, clustering and outlier detection, and molecular biology. The reverse $k$-nearest neighbor queries for a set of continuously moving objects has also attracted the attention of the database community; see~\cite{Cheema:2012:CRK:2124885.2124903} and references therein. Examples of moving objects include players in multi-player game environments, soldiers in a battlefield, tourists in dangerous environments, and mobile devices in wireless ad-hoc networks. 

To our knowledge, in computational geometry, there exist two data structures~\cite{MaheshwariVZ02cccg2002,CheongIJCGA2011} that give solutions to the \rknn~problem. Both of these solutions answer \rknn~queries for a set $P$ of stationary points and both only work for $k=1$. Maheshwari~\etal~(2002)~\cite{MaheshwariVZ02cccg2002} gave a data structure to solve the \r1nn~problem in $\mathbb{R}^2$. Their data structure creates an arrangement of largest empty circles centered at the points of $P$ and answers \r1nn~queries by point location in the arrangement. Their data structure uses $O(n)$ space and $O(n\log n)$ preprocessing time, and an \r1nn~query can be answered in time $O(\log n)$. Cheong~\etal~(2011)~\cite{CheongIJCGA2011} considered the \r1nn~problem in $\mathbb{R}^d$, where $d=O(1)$. Their method, which uses a compressed quadtree, partitions space into cells such that each cell contains a small number of candidate points. To answer an \r1nn query, their solution finds a cell that contains the query point and then checks all the points in the cell. Their approach uses $O(n)$ space and $O(n\log n)$ preprocessing time, and can answer an \r1nn~query in $O(\log n)$ time. It seems that the approach by Cheong~\etal~can be extended to answer \rknn~queries with preprocessing time $O(kn\log n)$, space $O(kn)$, and query time $O(\log n + k)$.

For a set $P$ of $n$ stationary points, one can report all the $1$-nearest neighbors in time $O(n\log n)$~\cite{Vaidya:1989:ONL:70530.70532}, and all the $k$-nearest neighbors, for any $k\geq 1$, in time $O(kn\log n)$~\cite{Dickerson:1996:APP:236464.236474}, where the neighbors are reported in order of increasing distance from each point; reporting the unordered set takes time $O(n\log n + kn)$~\cite{Callahan366854,Clarkson:1983:FAN:1382437.1382825,Dickerson:1996:APP:236464.236474}. 

For a set of moving points, there are three kinetic data structures (KDS's)~\cite{Agarwal:2008:KDD:1435375.1435379,DBLP:journals/corr/RahmatiA13,socg17-rahmati} to maintain all the $k$-nearest neighbors, but they only work for $k=1$.
\vspace{-10pt}
\paragraph{\textbf{Our contribution.}} For a set $P$ of $n$ continuously moving points in $\mathbb{R}^d$, where the trajectory of each point is a polynomial function of at most constant degree $s$, we provide a simple kinetic approach to answer \rknn~queries on the moving points. In fact, we provide the \textit{first} solution to the kinetic \rknn~problem for \textit{any} $k\geq 1$ in \textit{any} fixed dimension $d$. To answer an \rknn~query for a query point $q\notin P$ at any time $t$, we partition the $d$-dimensional space into a constant number of cones around $q$, and then among the points of $P$ in each cone, we examine the $k$ points having shortest projections on the cone axis. We obtain $O(k)$ candidate points for $q$ such that $q$ might be one of their $k$-nearest neighbors at time $t$. To check which if any of these candidate points is a reverse $k$-nearest neighbor of $q$, we maintain the $k^{th}$ nearest neighbor $p_k$ of each point $p\in P$ over time. By checking whether $|pq|\leq |pp_k|$ we can easily check whether a candidate point $p$ is one of the reverse $k$-nearest neighbors of $q$ at time $t$.

In the preprocessing step, we introduce a method for reporting all the $k$-nearest neighbors for all the points $p\in P$ in order of increasing distance from $p$. For $k=\Omega(\log^{d-1}n)$, both our method and the method of Dickerson and Eppstein~\cite{Dickerson:1996:APP:236464.236474} give the same complexity, but in our view, our method is simpler in practice. 

In order to answer \rknn~queries, our kinetic approach maintains all the $k$-nearest neighbors over time. This is the \textit{first} KDS for maintenance of all the $k$-nearest neighbors in $\mathbb{R}^d$, for any $k\geq 1$. Our KDS uses $O(n\log^{d+1} n + kn)$ space and $O(n\log^{d+1} n + kn\log n)$ preprocessing time, and processes $O(\phi(s,n) * n^2)$ events, each in amortized time $O(\log n)$. Here, $\phi(s,n)$ is the complexity of the $k$-level of a set of $n$ partially-defined polynomial functions, such that each pair of them intersects at most $s$ times. The current bounds on $\phi(s,n)$ are as follows~\cite{chanii2005,Chan:2008:LAC:1377676.1377691}.
\vspace{-5pt}
\[
\phi(s,n) = 
\begin{cases}
        {O(n^{3/2} \log n)},  & \text{for $s=2$};\\
        {O(n^{5/3} \text{poly}\log n)},  & \text{for $s=3$};\\
        {O(n^{31/18} \text{poly}\log n)},  & \text{for $s=4$};\\
        {O(n^{161/90-\delta})},  & \text{for $s=5$, for some constant $\delta>0$};\\
        {O(n^{2-1/2s-\delta_s})},  & \text{for odd $s$, for some constant $\delta_s>0$};\\
        {O(n^{2-1/2(s-1)-\delta_s})}, & \text{for even $s$, for some constant $\delta_s>0$}.
\end{cases}
\]
At any time $t$, an \rknn~query can be answered in time $O(\log^d n+k)$. Note that if an event occurs at the same time $t$, we first spend amortized time $O(\log n)$ to update all the $k$-nearest neighbors, and then we answer the query.
\vspace{-10pt}
\paragraph{\textbf{Outline.}} Section~\ref{sec:keyLemmas} provides two key lemmas, and in fact introduces a new supergraph, namely the \textit{$k$-Semi-Yao graph}, of the $k$-nearest neighbor graph. In Section~\ref{sec:ReportKNNs}, we show how to report all the $k$-nearest neighbors. Section~\ref{sec:KineticRkNNQueries} gives a (kinetic) data structure for answering  \rknn~queries on moving points, where the trajectory of each point is a bounded-degree polynomial. Section~\ref{sec:conclusion} concludes.
\section{Key Lemmas}\label{sec:keyLemmas}
\begin{figure}[t]
\begin{center}
\includegraphics[scale=1]{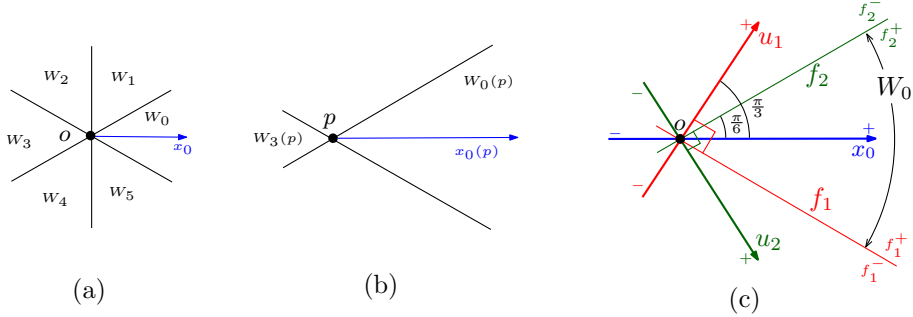}
\end{center}
\vspace{-20pt}
\caption{\small (a) A Partition of the plane into six wedges with common apex at $o$. (b) A translation of $W_0$ that moves apex to $p$. The wedge $W_0(p)$ is the reflection through $p$ of $W_3(p)$ and vise-versa. (c) The wedge $W_0$ in $\mathbb{R}^2$ is bounded by $f_1$ and $f_2$. The coordinate axes $u_1$ and $u_2$ are orthogonal to $f_1$ and $f_2$.}
\vspace{-15pt}
\label{fig:keyLemma}
\end{figure}

\vspace{-5pt}
Partition the plane around the origin $o$ into six wedges, $W_0,...,W_5$, each of angle $\pi/ 3$ (see Figure~\ref{fig:keyLemma}(a)). Denote by $W_l(p)$ the translation of wedge $W_l$, $0\leq l\leq 5$, such that its apex moves from $o$ to point $p$ (see Figure~\ref{fig:keyLemma}(b)). Denote by $x_l$ (resp. $x_l(p)$) the vector along the bisector of $W_l$ (resp. $W_l(p)$) directed outward from the apex at $o$ (resp. $p$). Denote the reflection of $W_l(p)$ through $p$ by $W_{l'}(p)$. Note that $l'=(l+3)\bmod {6}$; see Figure~\ref{fig:keyLemma}(b).
Consider the $i^{th}$ nearest neighbor $p_i$ of $p$. Denote by $L(P\cap W_l(p_i))$ the list of the points in $P\cap W_l(p_i)$, sorted by increasing order of their $x_l$-coordinates (projections). The following lemma provides a key insight. The short proof is omitted (see the full version of the paper in Chapter 6 of the first author's PhD dissertation~\cite{rahmati2014simple}).
\begin{lemma}\label{the:keyLemma2}
Let $p_i$ be the $i^{th}$ nearest neighbor of $p$ among a set $P$ of points in $\mathbb{R}^2$, and let $W_{l}(p_i)$ be the wedge of $p_i$ that contains $p$. Then point $p$ is among the first $i$ points in $L(P\cap W_l(p_i))$.
\end{lemma}
The \textit{$k$-nearest neighbor graph} (\knng) of a point set $P$ is constructed by connecting each point in $P$ to all its $k$-nearest neighbors. If we connect each point $p\in P$ to the first $k$ points in the sorted list $L(P\cap W_l(p))$, for $l=0,...,5$, we obtain what we call the \textit{$k$-Semi-Yao graph} (\ksyg). Lemma~\ref{the:keyLemma2} gives a necessary condition for $p_i$ to be the $i^{th}$ nearest neighbor of $p$: the point $p$ is among the first $i$ points in $L(P\cap W_l(p_i))$, where $l$ is such that $p\in W_l(p_i)$.  Therefore, the edge set of the \ksyg~covers the edges of the \knng. In summary, we have the following.
\begin{lemma}\label{the:NNGsubSYG}
The \knng~of a set $P$ of points in $\mathbb{R}^2$ is a subgraph of the \ksyg~of the set $P$.
\end{lemma}
\section{Reporting All $k$-Nearest Neighbors}\label{sec:ReportKNNs}
\vspace{-5pt}
Here we give a simple method for reporting all the $k$-nearest neighbors via a construction of the \ksyg. 

Let $C$ be a \textit{right circular cone} in $\mathbb{R}^d$ with opening angle $\theta$ with respect to some given unit vector $v$. Thus $C$ is the set of points $x\in \mathbb{R}^d$ such that the angle between $\overrightarrow{ox}$ and $\overrightarrow{v}$ is at most $\theta/2$. The angle between any two rays inside $C$ emanating from the apex $o$ is at most $\theta$. From now on, we assume $\theta\leq \pi/3$. 

Now consider a \textit{polyhedral cone} inscribed in the right circular cone $C$ where the polyhedral cone is formed by the intersection of $d$ distinct half-spaces, bounded by $f_1,...,f_d$, passing through the apex of $C$. Assuming $d$ is arbitrary but fixed, the $d$-dimensional space around the origin $o$ can be tiled by a constant number of polyhedral cones $W_0,...,W_{c-1}$~\cite{Abam:2011:KSX:1971362.1971367,Agarwal:2008:KDD:1435375.1435379}. Denote by $C_l$ the associated right circular cone of the polyhedral cone $W_l$. Let $x_l$ be the vector in the direction of the symmetry of $C_l$. Denote by $W_l(p)$ the translation of the wedge (polyhedral cone) $W_l$ where $o$ moves to $p$.

A similar approach and analysis as that in Section~\ref{sec:keyLemmas} can be easily used to state (key) Lemmas~\ref{the:keyLemma2} and~\ref{the:NNGsubSYG} for a set of points in $\mathbb{R}^d$.

To construct the \ksyg~efficiently, we need a data structure to perform the following operation efficiently: For each $p\in P$ and any of its wedges $W_l(p)$, $0\leq l\leq c-1$, find the first $k$ points in $L(P\cap W_l(p))$. Such an operation can be performed by using \textit{range tree} data structures. For each wedge $W_l$ with apex at origin $o$, we construct an associated $d$-dimensional range tree ${\cal T}_l$ as follows.

Consider a particular wedge $W_l$ with apex at $o$. The wedge $W_l$ is the intersection of $d$ half-spaces $f^+_1,...,f^+_d$ bounded  by $f_1,...,f_d$ (see Figure~\ref{fig:keyLemma}(c)). Let $\hat{u_j}$ denote the normal to $f_j$ pointing to $f^+_j$. We define $d$ coordinate axes $u_j$, $j=1,...,d$, through $\hat{u_j}$, where  $\hat{u_j}$ gives the respective directions of increasing $u_j$-coordinate values.

The range tree ${\cal T}_l$ is a regular $d$-dimensional range tree based on the $u_j$-coordinates, $j=1,...,d$. The points at level $j$ are sorted at the leaves according to their $u_j$-coordinates (for more details about range trees, see Chapter 5 of~\cite{Berg:2008:CGA:1370949}). Any $d$-dimensional range tree, \eg, ${\cal T}_l$, uses $O(n\log^{d-1} n)$ space and can be constructed in time $O(n\log^{d-1} n)$; for any point $r\in \mathbb{R}^d$, the points of $P$ inside the query wedge $W_l(r)$ whose sides are parallel to $f_j$, $j=1,...,d$, can be reported in time $O(\log^{d-1} n + z)$, where $z$ is the cardinality of the set $P\cap W_l(r)$~\cite{Berg:2008:CGA:1370949}. 

Now we add a new level to ${\cal T}_l$, based on the coordinate $x_l$. Let ${\cal C}_l(p)$ be the set of the first $k$ points in $L(P\cap W_l(p))$. To find ${\cal C}_l(p)$ in an efficient time, we use the level $d+1$ of ${\cal T}_l$, which is constructed as follows: For each internal node $v$ at level $d$ of ${\cal T}_l$, we create a list $L(P(v))$ sorted by increasing order of $x_l$-coordinates of the points in $P(v)$. For the set $P$ of $n$ points in $\mathbb{R}^d$, the range tree ${\cal T}_l$, which now is a $(d+1)$-dimensional range tree, uses $O(n\log^d n)$ space and can be constructed in time $O(n\log^d n)$.
 
The following lemma establishes the processing time for obtaining a ${\cal C}_l(p)$. The short proof is omitted (see the full version of the paper).
\begin{lemma}\label{the:SortingLists}
Given ${\cal T}_l$, the set ${\cal C}_l(p)$ can be found in time $O(\log^d n+k)$.
\end{lemma}
By Lemma~\ref{the:SortingLists}, we can efficiently find all the ${\cal C}_l(p)$, for all the points $p\in P$. This gives the following lemma.
\begin{lemma}\label{the:kSYG_Construction}
Using a data structure of size $O(n\log^d n)$, the edges of the $\ksyg$ of a set of $n$ points in fixed dimension $d$ can be reported in time $O(n\log^d n+kn)$.
\end{lemma}
Next, suppose we are given the \ksyg~and we want to report all the $k$-nearest neighbors. Let $E_p$ be the set of edges incident to the point $p$ in the \ksyg. By sorting these edges in non-decreasing order according to their Euclidean lengths, which can be done in time $O(|E_p|\log |E_p|)$, we can find the $k$-nearest neighbors of $p$  ordered by increasing distance from $p$. Since the number of edges in the $\ksyg$ is $O(kn)$ and each edge $pp'$ belongs to exactly two sets $E_p$ and $E_{p'}$, the time to find all the $k$-nearest neighbors, for all the points $p\in P$, is $\sum_{p} O(|E_p|\log |E_p|) = O(kn\log n)$. 

From the above discussion and Lemmas~\ref{the:NNGsubSYG} and~\ref{the:kSYG_Construction}, the following results.
\vspace{-10pt}
\begin{theorem}\label{the:kNNG_Construction}
For a set of $n$ points in fixed dimension $d$, our data structure can report all the $k$-nearest neighbors, in order of increasing distance from each point, in time $O(n\log^d n + kn\log n)$. The data structure uses $O(n\log^d n + kn)$ space.
\end{theorem}
\section{\rknn~Queries on Moving Points}\label{sec:KineticRkNNQueries}
\vspace{-5pt}
We are given a set $P$ of $n$ continuously moving points, where the trajectory of each point in $P$ is a polynomial function of bounded degree $s$. To answer \rknn~queries on the moving points, we must keep a valid range tree and track all the $k$-nearest neighbors during the motion. This section first shows how to maintain a (ranked-based) range tree, and then provides a KDS for maintenance of the \ksyg, which in fact gives a supergraph of the \knng~over time. Using the kinetic \ksyg, we can easily maintain all the $k$-nearest neighbors over time. Finally we show how to answer \rknn~queries on the moving points.
\paragraph{Kinetic RBRT.}\label{sec:KineticRBRT}
Let $u_j$, $1\leq j\leq d$, be the coordinate axis orthogonal to the half-space $f_j$ of the wedge $W_l$, $0\leq l\leq c-1$ (see Figure~\ref{fig:keyLemma}(c)). Abam and de Berg~\cite{Abam:2011:KSX:1971362.1971367} introduced a variant of the range tree, namely the \textit{ranked-based range tree} (RBRT), which has the following properties. Denote by ${\cal T}_l$ the RBRT corresponding to the wedge $W_l$.
\begin{itemize}
\item ${\cal T}_l$ can be described as a set of pairs $\Psi_l=\{(B_1,R_1),...,(B_m,R_m)\}$ such that:
\begin{itemize}
\item For any two points $p$ and $q$ in $P$ where $q\in W_l(p)$, there is a unique pair $(B_i,R_i)\in \Psi_l$ such that $p\in B_i$ and $q\in R_i$.
\item For any pair $(B_i,R_i)\in \Psi_l$, if $p\in B_i$ and $q\in R_i$, then $q\in W_l(p)$ and $p\in W_{l'}(q)$; here $W_{l'}(q)$ is the reflection of $W_l(q)$ through $q$.
\end{itemize}
The $\Psi_l$ is called a \textit{cone separated pair decomposition} (CSPD) for $P$ with respect to $W_l$. Each pair $(B_i,R_i)$ is generated from an internal node $v$ at level $d$ of the RBRT ${\cal T}_l$.
\item Each point $p\in P$ is in $O(\log^d n)$ pairs of $(B_i,R_i)$, which means that the number of elements of all the pairs $(R_i,B_i)$ is $O(n\log^d n)$. 
\item For any point $p\in P$, all the sets $B_i$ (resp. $R_i$) where $p\in B_i$ (resp. $p\in R_i$) can be found in time $O(\log^d n)$.
\item The set $P\cap W_l(p)$ is the union of $O(\log^d n)$ sets $R_i$, where $p\in B_i$.
\item When the points are moving, ${\cal T}_l$ remains unchanged as long as the order of the points along axes $u_j$, $1\leq j\leq d$, remains unchanged.
\item When a $u$-swap event occurs, meaning that two points exchange their $u_j$-order, the RBRT ${\cal T}_l$ can be updated in worst-case time $O(\log^d n)$ without rebalancing operations.
\end{itemize}
\subsection{Kinetic \ksyg}\label{sec:kineticKSYG}
\vspace{-5pt}
Here we give a KDS  for the \ksyg, for any $k\geq 1$, extending~\cite{DBLP:journals/corr/RahmatiA13}.

To maintain the \ksyg, we must track the set ${\cal C}_l(p)$ for each point $p\in P$. So, for each $1\leq i\leq m$, we need to maintain a sorted list $L(R_i)$ of the points in $R_i$ in ascending order according to their $x_l$-coordinates over time. Note that each set $R_i$ is some $P(v)$, the set of points at the leaves of the subtree rooted at some internal node $v$ at level $d$ of ${\cal T}_l$. To maintain these sorted lists $L(R_i)$, we add a new level to the RBRT ${\cal T}_l$; the points at the new level are sorted at the leaves in ascending order according to their $x_l$-coordinates. Therefore, in the modified RBRT ${\cal T}_l$, in addition to the $u$-swap events, we handle new events, called \textit{$x$-swap events}, when two points exchange their $x_l$-order. The modified RBRT ${\cal T}_l$ behaves like a $(d+1)$-dimensional RBRT. From the last property of an RBRT above, when a $u$-swap event or an $x$-swap event occurs, the RBRT ${\cal T}_l$ can be updated in worst-case time $O(\log^{d+1}n)$.

Denote by $\ddot{p}_{l,k}$ the $k^{th}$ point in $L(P\cap W_l(p))$.  To track the sets ${\cal C}_l(p)$, for all the points $p\in P$, we need to maintain the following over time.
\begin{itemize}
\item A set of $d+1$ \textit{kinetic sorted lists} $L_j(P)$, $j=1,...,d$, and the $L_l(P)$ of the point set $P$. We use these kinetic sorted lists to track the order of the points in the coordinates $u_j$ and $x_l$, respectively.
\item For each $B_i$, a sorted list $L(B'_i)$ of the points in $B'_i$, where $B'_i=\{(p,\ddot{p}_{l,k})|~p\in B_i\}$. The order of the points in $L(B'_i)$ is according to a \textit{label} of the second points $\ddot{p}_{l,k}$. This sorted list $L(B'_i)$ is used to answer the following query efficiently: Given a query point $q$ and a $B_i$, find all the points $p\in B_i$ such that $\ddot{p}_{l,k}=q$.
\item The $k^{th}$ point $r_{i,k}$ in the sorted list $L(R_i)$. We track the values $r_{i,k}$ in order to make necessary changes to the \ksyg~when an $x$-swap event occurs.
\end{itemize}
\vspace{-10pt}
\paragraph{Handling $u$-swap events.}
W.l.o.g., let $q\in W_l(p)$ before the event. When a $u$-swap event between $p$ and $q$ occurs, the point $q$ moves outside the wedge $W_l(p)$; after the event, $q\notin W_l(p)$. Note that the changes that occur in the \ksyg~are the deletions and insertions of the edges incident to $p$ inside the wedge $W_l(p)$. 

Whenever two points $p$ and $q$ exchange their $u_j$-order, we do the following updates.
\begin{itemize}
\item We update the kinetic sorted list $L_j(P)$. Each swap event in a kinetic sorted list can be handled in time $O(\log n)$.
\item We update the RBRT ${\cal T}_l$ and if a point is deleted or inserted into a $B_i$, we update the sorted list $L(B'_i)$. Since each insertion/deletion to $L(B'_i)$ takes $O(\log n)$ time, and since each point is in $O(\log^d n)$ sets $B_i$, this takes $O(\log^{d+1}n)$ time.
\item We update the values of $r_{i,k}$. After updating the RBRT ${\cal T}_l$, point $q$ might be inserted or deleted from some $R_i$ and change the values of $r_{i,k}$. So, for all $R_i$ where $q\in R_i$, before and after the event, we do the following. We check whether the $x_l$-coordinate of  $q$ is less than or equal to the $x_l$-coordinate of $r_{i,k}$; if so, we take the successor or predecessor point of $r_{i,k}$ in $L(R_i)$ as the new value for $r_{i,k}$. This takes $O(\log^{d+1} n)$ time.
\item We query to find ${\cal C}(p)$. By Lemma~\ref{the:SortingLists}, this takes $O(\log^d n +k)$ time.
\item If we get a new value for $\ddot{p}_{l,k}$, we update all the sorted lists $L(B'_i)$ such that $p\in B_i$. This takes $O(\log^{d+1} n)$ time.
\end{itemize}
Considering the complexity of each step above, and assuming the trajectory of each point is a  bounded degree polynomial, the following results.
\begin{lemma}\label{the:UswapEvents}
Our KDS for maintenance of the \ksyg~handles $O(n^2)$ $u$-swap events, each in worst-case time $O(\log^{d+1}n +k)$.
\end{lemma}
\vspace{-10pt}
\paragraph{Handling $x$-swap events.}
When an $x$-swap event between two consecutive points $p$ and $q$ with $p$ preceding $q$ occurs, it does not change the elements of the pairs $(B_i,R_i)$ of the CSPD $\Psi_l$. Such an event changes the $k$-SYG if both $p$ and $q$ are in the same $W_l(w)$, for some $w\in P$, and $w_{l,k}=p$. 

We apply the following updates to our KDS when two points $p$ and $q$ exchange their $x_l$-order.
\begin{enumerate}
\item We update the kinetic sorted list $L_l(P)$; this takes $O(\log n)$ time.
\item We update the RBRT ${\cal T}_l$, which takes $O(\log ^{d+1} n)$ time.
\item We find all the sets $R_i$ where both $p$ and $q$ belong to $R_i$ and such that $r_{i,k}=p$. Also, we find all the sets $R_i$ where $r_{i,k}=q$. This takes $O(\log^d n)$ time.
\item For each $R_i$, we extract all the pairs $(w,\ddot{w}_{l,k})$ from the sorted lists $L(B'_i)$ such that $\ddot{w}_{l,k}=p$. Note that each change to the pair $(w,\ddot{w}_{l,k})$ is a change to the \ksyg.
\item For each $w$, we update all the sorted lists $L(B'_i)$ where $(w,\ddot{w}_{l,k})\in B'_i$: we replace the previous value of $\ddot{w}_{l,k}$, which is $p$, by the new value $q$.
\end{enumerate}
Denote by $\chi_k$ the number of exact changes to the \ksyg~ of a set of moving points over time. For each found $R_i$, the fourth step takes $O(\log n + \xi_i)$ time, where $\xi_i$ is the number of pairs $(w,\ddot{w}_{l,k})$ such that $\ddot{w}_{l,k}=p$. For all these $O(\log^d n)$ sets $R_i$, this step takes $O(\log^{d+1}n +\sum_i \xi_i)$ time, where $\sum_i \xi_i$ is the number of exact changes to the \ksyg~when an $x$-swap event occurs. Therefore, for all the $O(n^2)$ $x$-swap events, the total processing time for this step is $O(n^2\log^{d+1}n+\chi_k)$.

The processing time for the fifth step  is a function of $\chi_k$. For each change to the \ksyg, this step spends $O(\log^{d+1}n)$ time to update the sorted lists $L(B'_i)$. Therefore, the total processing time for all the $x$-swap events in this step is $O(\chi_k*\log^{d+1}n)$.

From the above discussion and an upper bound for $\chi_k$ in Lemma~\ref{the:allKSYGchanges}, Lemma~\ref{the:XswapEvents} results. The proof of Lemma~\ref{the:allKSYGchanges} is omitted (see the full version of the paper).
\begin{lemma}\label{the:allKSYGchanges}
The number of changes to the \ksyg~of a set of $n$ moving points, where the trajectory of each point is a polynomial function of at most constant degree $s$, is $\chi_k=O(\phi(s,n)*n)$.
\end{lemma}
\begin{lemma}\label{the:XswapEvents}
Our KDS for maintenance of the \ksyg~handles $O(n^2)$ $x$-swap events with a total cost of $O(\phi(s,n)*n\log^{d+1}n)$.
\end{lemma}
From Lemmas~\ref{the:UswapEvents} and~\ref{the:XswapEvents}, the following theorem results.
\begin{theorem}\label{the:KinetickSYG}
For a set of $n$ moving points in $\mathbb{R}^d$, where the trajectory of each point is a polynomial function of at most constant degree $s$, our \ksyg~KDS uses $O(n\log^{d+1} n)$ space and handles $O(n^2)$ events with a total cost of $O(kn^2 + \phi(s,n)*n\log^{d+1}n)$.
\end{theorem}
\subsection{Kinetic All $k$-Nearest Neighbors}
\vspace{-5pt}
Given a KDS for maintenance of the \ksyg~(from Theorem~\ref{the:KinetickSYG}), a supergraph of the \knng, this section shows how to maintain all the $k$-nearest neighbors over time. For maintenance of the $k$-nearest neighbors of each point $p\in P$, we only need to track the order of the edges incident to $p$ in the \ksyg~according to their Euclidean lengths. This can easily be done by using a kinetic sorted list. The following theorem summarizes the complexity of our kinetic approach. The proof is omitted (see the full version of the paper).
\begin{theorem}\label{the:KinetickNNs}
For a set of $n$ moving points in $\mathbb{R}^d$, where the trajectory of each point is a polynomial of at most constant degree $s$, our KDS for maintenance of all the $k$-nearest neighbors, ordered by distance from each point, uses $O(n\log^{d+1} n +kn)$ space and $O(n\log^{d+1} n + kn\log n)$ preprocessing time. Our KDS handles $O(\phi(s,n) * n^2)$ events, each in $O(\log n)$ amortized time. 
\end{theorem}
\subsection{\rknn~Queries}
\vspace{-5pt}
Suppose we are given a query point $q\notin P$ at some time $t$. To find the reverse $k$-nearest neighbors of $q$, we seek the points in $P\cap W_l(q)$ and find ${\cal C}_l(q)$, the set of the first $k$ points in $L(P\cap W_l(q))$. The set $\cup_l{\cal C}_l(q)$ contains $O(k)$ candidate points for $q$ such that $q$ might be one of their $k$-nearest neighbors. In time $O(\log^d n)$ we can find a set of $R_i$ where $P\cap W_l(q)=\sum_i R_i$. From Lemma~\ref{the:SortingLists}, and since we have sorted lists $L(R_i)$ at level $d+1$ of ${\cal T}_l$, the $O(k)$ candidate points for the query point $q$ can be found in worst-case time $O(\log^d n +k)$. Now we check whether these candidate points are the reverse $k$-nearest neighbors of the query point $q$ at time $t$ or not; this can be easily done by application of Theorem~\ref{the:KinetickNNs}, which in fact maintain the $k^{th}$ nearest neighbor $p_k$ of each $p\in P$. Therefore, checking a candidate point can be done in $O(1)$ time by comparing distance $|pq|$ to distance $|pp_k|$. This implies that checking which elements of ${\cal C}_l(q)$, for $l=0,...,c-1$, are reverse $k$-nearest neighbors of the query point $q$ takes time $O(k)$.

If a query arrives at a time $t$ that is simultaneous with the time when one of the $O(\phi(s,n) * n^2)$ events occurs, our KDS first spends amortized time $O(\log n)$ to handle the event, and then spends time $O(\log^d n +k)$ to answer the query. Thus we have the following.
\begin{theorem}\label{the:KineticRkNNQ}
Consider a set $P$ of $n$ moving points in $\mathbb{R}^d$, where the trajectory of each one is a bounded-degree polynomial. The number of reverse $k$-nearest neighbors for a query point $q\notin P$ is $O(k)$. Our KDS uses $O(n\log^{d+1} n + kn)$ space, $O(n\log^{d+1} n + kn\log n)$ preprocessing time, and handles $O(\phi(s,n)*n^2)$ events. At any time $t$, an \rknn~query can be answered in time $O(\log^d n+k)$. If an event occurs at time $t$, the KDS spends amortized time $O(\log n)$ on updating itself.
\end{theorem}
\section{Discussion}\label{sec:conclusion}
\vspace{-5pt}
In the kinetic setting, where the trajectories of the points are polynomials of bounded degree, to answer the \rknn~queries over time we have provided a KDS for maintenance of all the $k$-nearest neighbors. Our KDS is the first KDS for maintenance of all the $k$-nearest neighbors in $\mathbb{R}^d$, for any $k\geq 1$. It processes $O(\phi(s,n) * n^2)$ events, each in amortized time $O(\log n)$. An open problem is to design a KDS for all $k$-nearest neighbors that processes less than $O(\phi(s,n) * n^2)$ events.

Arya~\etal~\cite{Arya:1998:OAA:293347.293348} have a kd-tree implementation to approximate the nearest neighbors of a query point that is in use by practitioners~\cite{10.1109/TVCG.2010.9} who have found challenging to implement the theoretical algorithms~\cite{Callahan366854,Clarkson:1983:FAN:1382437.1382825,Dickerson:1996:APP:236464.236474,Vaidya:1989:ONL:70530.70532}. Since to report all the $k$-nearest neighbors ordered by distance from each point our method uses multidimensional range trees, which can be easily implemented, we believe our method may be useful in practice.
\subsubsection*{Acknowledgments.}{We thank Timothy M. Chan for his helpful comments and suggestions.}
\bibliographystyle{splncs03}
\bibliography{References_Main}
\end{document}